\documentclass[12pt]{article}
\usepackage{amsmath,amssymb,amsfonts}
\usepackage[paper=letterpaper,margin=1.0in]{geometry}
\usepackage{graphicx}

\newcommand{\vev}[1]{\langle #1 \rangle}
\newcommand{\bvev}[1]{\bigl\langle #1 \bigr\rangle}

\newcommand{\dvev}[1]{\langle\!\langle #1 \rangle\!\rangle}
\newcommand{\bdvev}[1]{\bigl\langle\!\bigl\langle #1 \bigl\rangle\!\bigr\rangle}

\newcommand{\btvev}[1]{\bigl\langle\!\bigl\langle\!\bigl\langle #1 \bigl\rangle\!\bigl\rangle\!\bigr\rangle}

\newcommand{\Ahat}{\hat{A}}
\newcommand{\Bhat}{\hat{B}}

\renewcommand{\thefootnote}{\fnsymbol{footnote}}
\begin{document}

\rightline{\footnotesize IPMU13-0065}

\begin{center}

\vskip 1.0cm
\centerline{\Large\bf 
Is Quantum Gravity a Super-Quantum Theory?
}
\vskip 1.0cm

\centerline{\bf
Lay Nam Chang${}^{1}$\footnote{laynam@vt.edu},
Zachary Lewis${}^{1}$\footnote{zlewis@vt.edu},
Djordje Minic${}^{1}$\footnote{dminic@vt.edu} and
Tatsu Takeuchi${}^{1,2}$\footnote{takeuchi@vt.edu}
}

\vskip 0.5cm

{\it
${}^1$Department of Physics MC 0435, Virginia Tech, Blacksburg, VA 24061, USA \\
\vskip 0.2cm
${}^2$Kavli Institute for the Physics and Mathematics of the Universe (WPI),\\
The University of Tokyo, Kashiwa-shi, Chiba-ken 277-8583, Japan
}

\vskip 1.0cm
\centerline{Submission date: March 26, 2013}
\vskip 1.0cm

\begin{abstract}
We argue that quantum gravity should be a super-quantum theory, that is, a theory
whose non-local correlations are stronger than those of canonical quantum theory.
As a super-quantum theory, quantum gravity should display distinct experimentally observable 
super-correlations of entangled stringy states. 
\end{abstract}

\vspace{1cm}

\centerline{Essay written for the Gravity Research Foundation 2013 Awards for Essays on Gravitation}

\end{center}

\setcounter{footnote}{0}
\renewcommand{\thefootnote}{\arabic{footnote}}
\newpage

\noindent
\textbf{\large 1. Three Approaches to Quantum Gravity}
\medskip

Any quantum theory of gravity must necessarily be a 
theory which encompasses special relativity, quantum mechanics (QM), and Newtonian 
Gravity. Representing these three essential ingredients with the three 
fundamental constants $c$, $\hbar$, and $G_N$, 
we can classify the approaches to quantum gravity according to the ordering in 
which these constants are incorporated.

In 
the first approach, $c$ and $\hbar$ are incorporated into relativistic 
quantum field theory (QFT), to which string theory adds $G_N$ 
via the introduction of the length scale $\ell_s = \ell_P = \sqrt{G_N\hbar/c^3}$ \cite{string}. 
String theory is a natural extension of QFT to an 
infinite number of fields. One of its massless excitations is 
the spin-2 graviton so features of general relativity (GR) 
emerge at long distances. 
However, despite recent 
advances in non-perturbative string theory, its basic principles 
are yet to be understood.  
The issues of 
background independence, the nature of the Big Bang singularity, the 
vacuum energy problem, as well as the origins of the 
Standard Model of particle physics and of dark matter remain 
unresolved. One thing is certain in this approach: gravity is 
emergent and not fundamental, and thus spacetime is as well.  
The nature of the underlying quantum degrees of freedom from 
which spacetime and gravity must emerge is still mysterious.

The 
second approach starts from GR, which unifies $c$ and $G_N$, 
and attempts to quantize its dynamical degrees of freedom, \textit{i.e.} 
spacetime itself. Perhaps the best example of this approach is 
given by loop quantum gravity \cite{loop}.  In this approach the geometric 
nature of gravity is taken into account from the beginning. 
The resulting structure exploits the properties of non-local loop variables, 
yet is radically different from string theory, as string theory 
also features higher dimensionality of spacetime, supersymmetry, and supergravity. Loop 
gravity does not require these additional structures and essentially proposes 
a discrete fundamental structure for spacetime that is consistent with 
Lorentz symmetry and background independence. In this approach, matter and 
gravity do not go hand in hand as in string 
theory, and the origin of the matter sector is usually 
not considered to be one of its central questions.

The 
third avenue starts from a theory defined by $\hbar$ and 
$G_N$, a non-relativistic quantum theory of gravity \cite{MLUR}.  One version of 
this unconventional, yet logically viable, option was advocated by us 
in Ref.~\cite{chia}, and begins with Matrix theory, a version 
of light-cone string field theory at infinite coupling, and also 
a natural regularization of the theory of membranes, viewed as 
solitonic objects in a strongly coupled string theory of a 
particular type \cite{matrix}. Though Matrix theory does not look like 
a theory of gravity, being a supersymmetric Matrix QM with 
a non-Abelian gauge structure, it reproduces Newton's law of gravitation 
at long distances, and attains relativistic covariance in a limit 
defined by an infinite number of its quanta. To Matrix 
theory, we add a new ingredient: in order to address 
the question of background independence, we argue that the canonical 
geometric structure of QM, represented by complex projective spaces, must 
be made dynamical and generalized to the non-linear Grassmannians \cite{mm}. Essentially, 
our proposal is a generalized geometric Matrix QM, which takes 
the central lessons of the first two approaches quite seriously. 
It uses Matrix theory, which appears in the context of 
non-perturbative string theory, and it emphasizes geometry and background independence, 
which are the hallmarks of loop quantum gravity.

In all 
three approaches, the attempts to incorporate the third constant into 
the formalisms suggest that fundamental departures from the conventional QM 
framework is inevitable for quantum gravity. What is the new 
framework of physics that quantum gravity demands?  This new framework 
should not only deepen our understanding of gravity, but 
also shed light on the mysteries of 
QM, just as QM had shed light on the structure 
of classical mechanics.  
Independently of what this framework may be,
we argue 
that it would be characterized by correlations which are stronger 
than those of QM in the context of Bell's inequalities \cite{bell}.

\bigskip
\noindent
\textbf{\large 2. Super-Quantum Correlations and Quantum Gravity}
\medskip

Here, we adopt the version of Bell's inequality as formulated
in Ref.~\cite{CHSH}.  Let $A$ and $B$ represent the outcomes 
of measurements performed on some isolated physical system by detectors 
1 and 2 which are placed at two causally disconnected 
spacetime locations.  Assume that the only possible values of $A$ 
and $B$ are $\pm 1$.  Let $P(a,b)=\vev{A(a)B(b)}$ be the expectation value 
of the product $A(a)B(b)$ where $a$ and $b$ respectively denote 
the settings of detectors 1 and 2. Then, the upper 
bound, $X$, of the following combination of correlators, for arbitrary 
detector settings {$a$, $a'$, $b$, $b'$}, characterizes each underlying theory:
\begin{equation}
\Bigl|
 P(a, b) 
+P(a, b')
+P(a', b) 
-P(a', b') 
\Bigr| \;\leq\; X\;.
\label{CHSHbound}
\end{equation}
This bound for classical hidden variable theories is $X_\mathrm{Bell}=2$, while 
that for QM is $X_\mathrm{QM}=2\sqrt{2}$. That is, QM correlations violate 
the classical Bell bound but are themselves bounded \cite{cirelson}.  The maximum 
possible value of $X$ is 4, and the requirement of 
relativistic causality does not preclude correlations which saturate this absolute 
bound \cite{super}.

What type of theory would predict such super-quantum correlations?
Since the process of quantization increases the bound from $X_\mathrm{Bell}=2$
to $X_\mathrm{QM}=2\sqrt{2}$, we proposed in Ref.~\cite{vtbell} the naive conjecture 
that another step of ``quantization'' would further increase the bound 
by a factor of $\sqrt{2}$ to 4.

What procedure would 
such a ``double'' quantization entail?  Quantization demands that correlation functions 
of operators be calculated via the path integral
\begin{equation}
\bvev{\,\Ahat(a)\Bhat(b)\,}  \;=\; \int \!Dx\;A(a,x)\,B(b,x)\,\exp\left[\dfrac{i}{\hbar} S(x)\right]
\;\equiv\;
A(a)\star B(b)
\;,
\label{AstarB}
\end{equation}
where $x$ 
collectively denotes the classical dynamical variables of the system. In 
a similar fashion, we can envision performing another step of 
quantization by integrating over ``paths'' of quantum operators to define 
correlators of ``super'' quantum operators
\begin{equation}
\bdvev{\,\hat\Ahat(a)\hat\Bhat(b)\,}  
\;=\; \int \!D\hat{\phi}\;\Ahat(a,\hat{\phi})\,\Bhat(b,\hat{\phi})\,\exp\left[\dfrac{i}{\tilde{\hbar}} \tilde{S}(\hat{\phi})\right]\;,
\end{equation}
where $\hat{\phi}$ collectively denotes the 
dynamical quantum operators of the system. Here, $\dvev{\hat\Ahat(a)\hat\Bhat(b)}$ is an 
operator.  To further reduce it to a number, we must 
calculate its expectation value in the usual way
\begin{equation}
\bdvev{\,\hat\Ahat(a)\hat\Bhat(b)\,}  
\quad\rightarrow\quad
\btvev{\,\hat\Ahat(a)\hat\Bhat(b)\,}  
\;=\;
\left\langle
\int \!D\hat{\phi}\;\Ahat(a,\hat{\phi})\,\Bhat(b,\hat{\phi})\,\exp\left[\dfrac{i}{\tilde{\hbar}} \tilde{S}(\hat{\phi})\right]
\right\rangle
\;,
\end{equation}
which would 
amount to replacing all the products of operators on the 
right-hand side with their first-quantized expectation values, or 
equivalently, replacing the operators with `classical' variables except with their 
products defined via Eq.~(\ref{AstarB}). Note that this is precisely 
the formalism of Witten's open string field theory (OSFT) \cite{witten}, in 
which the action for the `classical' open string field $\Phi$
is given formally as
\begin{equation}
S_W (\Phi) \;=\; \int \Phi \star Q_\mathrm{BRST} \Phi + \Phi \star \Phi \star \Phi\;,
\end{equation}
where $Q_\mathrm{BRST}$ is the open string 
theory BRST cohomology operator ($Q_\mathrm{BRST}^2=0$), and the star product is 
defined via a world-sheet path integral weighted with the Polyakov 
action and deformation parameter $\alpha'=\ell_s^2$.
The fully quantum OSFT is 
then, in principle, defined by yet another path integral in 
the infinite dimensional space of the open string field $\Phi$, 
i.e. 
\begin{equation}
\int D \Phi\,\exp\left[\dfrac{i}{g_s}S_W(\Phi)\right]\;,
\end{equation} 
where $g_s$ is the string coupling and all products 
are defined via the star-product. For reasons of unitarity, 
OSFT must contain closed strings, and therefore gravity. Thus, OSFT 
is a manifestly ``doubly'' quantized theory, and we argue that 
it, and the theory of quantum gravity it should become, 
would be characterized by super-quantum correlations when fully formulated.

Further insight can be obtained from toy models, such as 
those in Refs.~\cite{galois} and \cite{biorthogonal}. Specifically, super-quantum correlations 
were found in Ref.~\cite{biorthogonal}. There, the probabilities of individual 
outcomes were indeterminate while the expectation values of observables were 
determinable. This suggests that super-quantum correlations would result from 
a theory in which probability distributions themselves are probabilistically determined, 
again pointing to a ``double'' quantization.

``Double" quantization can also 
be qualitatively seen in loop quantum gravity, as spacetime itself 
is quantized before QFT is quantized on top of it. 
It is also likely present in the previously discussed geometric 
generalization of Matrix QM, as the additional quantum dynamics of 
the state space should represent a ``second" quantization. Though super-correlations 
are not (currently) clearly present in these contexts, we 
expect that they should occur. If so, it would add 
further support to the notion that super-correlations are a 
generic aspect of quantum gravity.

\newpage
\noindent
\textbf{\large 3. Experimental Signatures?} 
\medskip

In closing, let us offer 
some comments on possible experimental observations of such super-quantum 
violations of Bell's inequalities in quantum gravity.

The usual experimental 
setup for testing the violation of Bell's inequalities in QM 
involves entangled photons \cite{exp}. In OSFT, photons are the lowest lying 
massless states, but there is a whole Regge trajectory associated 
with them. The obvious experimental suggestion is to look for 
effects from entangled Reggeized photons. Such experiments are of course 
impossible at present, given their Planckian nature.

A more feasible 
place to look for super-quantum correlations could be in 
cosmological data. It is believed that quantum fluctuations seed the 
large scale structure of the Universe, i.e. galaxies and clusters 
of galaxies that we observe \cite{Bardeen:1983qw}. The simplest models use Gaussian 
quantum correlations, though non-Gaussian correlations are envisioned as well 
and are constrained by new data on the cosmic microwave 
background (CMB) from the Planck satellite \cite{planck}. 
While it is yet 
unclear how super-quantum correlations would affect the CMB data, 
we expect that they would leave ``stringy'' imprints on the 
large scale structure of the Universe and be observable at 
those scales.

\vskip 0.5cm

\newpage

\noindent
\textbf{\large Acknowledgements}
\medskip

ZL, DM, and TT were supported by the U.S. Department of Energy under contract DE-FG05-92ER40677, Task A. 
TT was also supported by the World Premier International Research Center Initiative (WPI Initiative), MEXT, Japan.


\end{document}